\documentstyle{article}
\begin{document}
\large
\baselineskip=24pt
\title{{\bf Domain Growth, Wetting and Scaling\\ in Porous Media}}
\author{D. W. Grunau\thanks{Department of Mathematics, Colorado
State University, Ft. Collins, CO 80523}, T. Lookman\thanks{Department
of Applied Mathematics, University of Western Ontario, London, Ontario
N6A5B7 CANADA}, S. Y. Chen, and A. S. Lapedes\\
  [.5cm]
        {\em Theoretical Division}\\
        {\em and}\\
        {\em Center for Non-linear Studies}\\
        {\em Los Alamos National Laboratory}\\
        {\em Los Alamos, NM 87545}\\[.4cm]}
\date{}
\maketitle

\vspace{0.4in}
\begin{center}
{\bf Abstract}\par
\end{center}

The lattice Boltzmann (LB) method is used to study the kinetics of domain
growth of a binary fluid in a number of geometries modeling porous media.
Unlike the traditional methods which solve the Cahn-Hilliard equation, the LB
method correctly simulates fluid properties, phase segregation, interface
dynamics and wetting. Our results, based on lattice sizes of up to $4096
\times 4096$, do not show evidence to indicate the breakdown of late stage
dynamical scaling, and suggest that confinement of the fluid is the key
to the slow kinetics observed. Randomness of the pore structure
appears unnecessary.

\newpage
It is well established that a binary fluid mixture undergoes phase
separation if rapidly quenched from a high temperature phase to a point in the
coexistence region. Moreover, when the size of the domains is much
larger than the interfacial thickness, there is only one dominant
length scale in the system \cite{rev}. This course of phase separation is
completely changed when the binary fluid mixture is contained within a
porous medium. Phase separation does not proceed to a macroscopic
scale, and many small domains of the two phases are formed, even far
below the critical point [2-4]. This lack of phase separation is poorly
understood, and whether the observed small domain structure is due to
the randomness of the pore structure or to the confinement of the
mixture in small pores is controversial. One suggestion is that the
randomness of the pore structure causes random-field Ising-like
behavior, which gives rise to the small domains, leading to the
metastability and the slow kinetics of domain growth that are observed
in experiments \cite{deg}. An alternate suggestion is that confinement slows
down domain growth \cite{liu}. A  Monte Carlo study of
the Ising model confined in a regular pore geometry allows for various
long-lived metastable configurations (`plugs',`capsules' and `tubes')
depending on temperature and strength of interaction, without any
randomness included \cite{liu1}. Recent work, using a Cahn-Hilliard
description, investigated the kinetics of domain growth in an
interconnected geometry resembling Vycor glass \cite{cha}. It concluded
that the interconnected and tortuous geometry creates barriers to domain
growth, and that there is breakdown of dynamical scaling at the late stage,
implying the presence of many length scales. Moreover, since slow
kinetics were obtained regardless of any surface interaction, the study
inferred that it was unnecessary to invoke the random field model.

The above research has stimulated our interest in studying domain growth
and wetting in porous media using a lattice Boltzmann method of simulation.
We argue that this technique
is a more appropriate tool to study the problem than the use of the
Cahn-Hilliard equation, or Ising type models. A number of geometries
with imposed boundary conditions such as non-wetting, wetting, slip
and no-slip conditions were investigated for lattice sizes of up to
$4096 \times 4096$.  We conclude that the evidence is not consistent
with the breakdown of dynamical scaling, in the late time regime.
In addition, we observe a slowdown of growth kinetics, regardless of
the shape of the medium or conditions at the walls.  This implies that,
for times accessible to simulation, confinement of the fluid
appears to be the key to the slowdown of kinetics, and that randomness
of the pore geometry is not necessary.

Lattice gas and lattice Boltzmann methods provide computational
environments with which to study hydrodynamic phase segregation
using parallel computing techniques [9-14]. They simulate fluid properties,
phase segregation and interface dynamics simultaneously, while
complex geometries and their associated boundary conditions are
implemented with simplicity. Other methods, such as molecular dynamics
simulations, can accurately represent the dynamics of real fluids, but
they are computationally intensive, limited to small systems, and are
unable to access the late regime of spinodal decomposition. Methods that
include hydrodynamic interactions, such as cell dynamical systems
\cite{shi}, and time dependent Ginzburg-Landau equations \cite{jef},
approximate the dynamics with Langevin equations which are then solved
numerically. In addition to being computationally intensive, these methods
sacrifice the Navier-Stokes behavior of the fluid and the interface
dynamics. On the other hand, LB methods simulate hydrodynamic phase
segregation in a natural way, without the introduction of {\em ad hoc}
relations between the order parameter fluctuations and the fluid dynamics.

The lattice Boltzmann method is a discrete in time and space, microscopic,
kinetic description of the evolution of a {\em particle distribution
function} of a fluid. Point particles move along the links of a lattice
(hexagonal in two dimensions), obey certain collision rules, and
macroscopically mimic the Navier-Stokes equations in certain limits
\cite{ccmm}. It has been used recently to study the effects of
hydrodynamics on the late stage kinetics of a binary fluid mixture
undergoing spinodal decomposition \cite{rothman,alx}. The details of
the steps involved in simulating the microscopic dynamics, such as streaming,
collision, interface perturbation and recoloring, as well as
comparison with experiment, are discussed in \cite{gru}.
In contrast to the lattice Boltzmann method, the Cahn-Hilliard
equation is more suitable for the study of binary alloys or glasses,
rather than flow through porous media \cite{rev}. This is due to its
inability to take into account hydrodynamic effects, which are important
in the late stages of spinodal decomposition. Moreover, it assumes
that the domain growth is the same, irrespective of which of the two
fluids wets the surface. This assumption is valid only in the absence
of wetting, the presence of which introduces an asymmetry such that
the Ginzburg-Landau free energy is no longer symmetric in the order
parameter. The Cahn-Hilliard equation, therefore, does not correctly
model wetting, which may explain some of the differences between our
results and those found in reference \cite{cha}. These difficulties are
overcome in our use of the LB method which correctly simulates
fluid-wall interfacial dynamics. Wetting is modeled by introducing a
colored-particle interaction between walls and their nearest neighbor
sites. This forces the wetting fluid in the porous medium to be attracted
to the walls \cite{gru}.

The two dimensional geometries that were studied included $(a)$ a random
distribution of disks with average radius 12 units, variance 4 units
and porosity of 95\%, $(b)$ an interconnected and tortuous geometry with
average
domain size of 160 units and porosity 75\% \cite{cha} and $(c)$ periodic
slits with pore radius of 16 and 64 lattice units. The geometry in $(b)$
allowed a close comparison of results by different methods. In
addition to studying various geometries, we also investigate the
effects of both slip and no-slip conditions at the walls. The no-slip
condition is accomplished by requiring a ``bounce back'' of particles
when they collide with a wall boundary, whereas the slip condition
invokes a specular reflection of such particles.  To study domain
growth sizes, we define a local order parameter as the density difference
between binary fluids, which we label red and blue:
$$\psi({\bf x},t)= \rho_r - \rho_b = \sum_{i=0}^{6}
(f_{i}^{r}({\bf x},t) - f_{i}^{b}({\bf x},t)).$$
Here, $f_{i}^{r}({\bf x},t)$ and $f_{i}^{b}({\bf x},t)$ are respectively
the particle distribution functions for red (r) and blue (b) fluids
at site ${\bf x}$ and time $t$ moving along a link having direction $i$,
and $\rho_r$ and $\rho_b$ are the red and blue fluid density. Also,
$f_{i}({\bf x},t)=f_{i}^{r}({\bf x},t)+f_{i}^{b}({\bf x},t)$ is the
particle distribution function for the total fluid, where $i=0,\ldots,6$
represents the velocity directions at each site of a hexagonal lattice.
The state $i=0$ corresponds to the portion of the fluid at rest.
The LB equation for $f_{i}({\bf x},t)$ can be written as
$$f_{i}^{k}({\bf x}+{\bf e}_{i},t+1)=
f_{i}^{k}({\bf x},t)+\Omega_{i}^{k},$$
where $k$ denotes a red or blue fluid species, and $(\Omega_{i}^{k})=
(\Omega_{i}^{k})^{c} + (\Omega_{i}^{k})^{p}$ is the collision
operator consisting of a term representing the change in
$f_{i}^{k}$ due to collisions ($\Omega_{i}^{k})^{c}$ and a term
representing the color perturbation ($\Omega_{i}^{k})^{p}$.
The vectors ${\bf e}_{i}$ are the velocity vectors along the
links of the lattice. The collision term $(\Omega_{i}^{k})^{c}$ is
chosen to have a single time relaxation form
$$(\Omega_{i}^{k})^{c}=-\frac{1}{\tau}(f_{i}^{(k)}-f_{i}^{(eq)}),$$
where $\tau$ is the characteristic relaxation time and
$f_{i}^{(eq)}$ is the local equilibrium distribution function \cite{ccmm,qian}.
The surface tension inducing perturbation $(\Omega_{i}^{k})^{p}$ and
the recoloring procedure are chosen appropriately so that Laplaces's
law holds for the model \cite{gru}. The mean free path in LB simulations
is of the order of one lattice unit. The geometries we used have a pore
radius of at least 16 units (greater than 50 for $4096 \times 4096$),
therefore, finite size effects associated with the pore sizes should be
insignificant.

Simulations were carried out on lattices of sizes up to $4096 \times 4096$,
for deep critical quenches with $\sum_{{\bf x}} \psi({\bf x},t) =0$.
The lattice was initialized with a random distribution of
red and blue fluids in the pores, in order to introduce an initial
fluctuation in the order parameter. Growth kinetics are characterized
through the order parameter correlation function
$$G(r,t) =<\psi(r)\psi(0)>-<\psi>^2,$$
where $\langle \rangle$ denotes a spatial average. A characteristic
domain size, $R(t)$, can be defined from $G(r,t)$ by its first zero
crossing, and a structure factor can be defined from $G(r,t)$ by its
Fourier transform. Note that the order
parameter $\psi$ is assigned a value of zero at the walls of the porous medium.

The domain morphology, with and without wetting, is shown in Fig. 1 for
two geometries: the random distribution of disks of average radius 12
and variance 4 lattice units (geometry $(a)$ defined above), and the
tortuous geometry used in \cite{cha} (geometry $(b)$ above).  Elongated
domains with a layer of wetting fluid are evident in Fig. 1a, whereas
the characteristic plugs, cusps, and straight interfaces of the non-wet
condition is seen in Fig. 1b.
The variation of domain size as a function of time for the tortuous geometry
of Fig. 1b is shown in Fig. 2a. During the early stages of growth
($t < 100$), where the two fluids are not completely separated, the
system is unaware of the walls and therefore the growth for the bulk
fluid case, non-wet walls, and wet walls show the same behavior.
The domain size, $R(t)$, shows the expected $t^{1/2}$ law for early
time \cite{rev}. As the dynamics evolves, however, the bulk fluid case
assumes the expected $R(t) \sim t^{2/3}$ form ({\small $\bigcirc$} signs)
for the late stage period of time, while the presence of the walls begin
to affect the domain growth for the other two cases ($\Diamond$ and
$+$ signs). It is interesting to note that in all of the porous media that
we studied, domains grew, for late time, as $t^{2/3}$ until their dynamics
were strongly affected by the wall geometry.
We expect that this period of time may
vary, depending upon the geometry, and may not exist at all for
geometries with a large degree of confinement.  Fig. 2a shows that late
times for the bulk fluid case are in the range of $1000<t<2000$ where the
domain size is $\sim$ 40-50 (much smaller than 4096).  The slowdown is
appreciable when the domain size is comparable to the characteristic size
of the porous medium ($--$), defined by the first zero crossing of the
pore-wall correlation function. When a fluid wets the walls, the domain
growth ($+$ signs) is not as slow as the non-wetting case, and can exceed the
characteristic pore size.  The large system size considered here ($4096\times
4096$) did not allow us to access late enough times to observe this, however,
it is clearly seen in Fig. 2b ($1024 \times 1024$).  Wetting effects
allow the elongation of this phase, as well as making it possible to form
a thin film surrounding the rock, increasing the period of time for domain
growth.
To investigate the effects of velocity boundary conditions on growth dynamics,
we compare, in Figure 2b, the domain growth for a non-wet, slip boundary
condition ({\small $\bigcirc$} signs), a non-wet, non-slip boundary condition
($\Diamond$ signs), and a wet, non-slip boundary condition ($+$ signs).
We find that the domain growth of both the non-wet, slip and the wet,
non-slip conditions exceeds the characteristic pore size ($--$) of
the given geometry.  It is of particular interest that the slip
condition facilitates domain growth to a greater degree than in the wetting,
no-slip case.  This illustrates that confinement may be of more
importance than wetting in the dynamics of domain growth slowdown.

The most interesting growth dynamics that we observed occur during the
time interval between the $t^{2/3}$ growth region ($t_i$) and
before phase segregation stops ($t_f$).  Note that since we do not stir
the fluids, the domain growth will ultimately cease.  During $t_i<t<t_f$,
there is a competing mechanism between the confinement of the
geometry and the inertial motion of the separated domains with
a curvature driven force \cite{rev}.  The former will
attempt to slow the fluid motion at the walls, whereas the latter will
try to keep the separated fluid domains in motion, allowing their
aggregation to continue. Note that the non-slip condition confines
fluid motion in both the normal and tangential directions, but the
slip condition has confinement only in the normal direction. Clearly
at this stage, the growth of the domain size will be slower than the
bulk fluid case. An interesting phenomenon is seen in Fig. 2c where
the domain growth is shown for both red and blue fluids in the presence
of wetting. During this stage, the domain growth is not the same for the
two fluids when the surface is preferentially wet for one fluid ($\Box$ signs)
while not for the other ({\small $\bigcirc$}).  On the other hand, when no
wettability is present, domain growth for both the red and blue fluid
is identical ($\triangle$).
This emphasizes that the Cahn-Hilliard approach, which uses only the order
parameter, does not model wetting correctly, and may not be an appropriate
model for the phase separation process when wetting is important. Moreover,
the behavior of domain growth in our simulations is different from those
shown in \cite{cha}, using the Cahn-Hilliard approach.  In our
results, the domain growth curves, in all cases, overlap for early
times, and then deviate as the dynamics evolves. The results of \cite{cha},
are, in fact, difficult to reconcile with the
expected physical effects which are correctly modeled by the LB approximation
in this paper.

The remaining question of importance, to be answered by our model, is whether
or not domain growth in the late-time region ($t_i<t<t_f$) breaks the
scaling laws as demonstrated by the previous study \cite{cha}. To answer
this question, we examine the structure factor, $S(k,t)$, which is known
to obey the dynamical scaling relation
$$S(k,t)=R(t)^dF(y)$$
for late times, in a wide variety of segregating systems \cite{rev}. Here
$F(y)$ is the scaling function, with $y=kR(t)$, and
$d$ is the spatial dimension. The function $F(y)$ is plotted versus
$y$ in Fig. 3 for the bulk fluid and non-wet boundary condition
(geometry $(b)$ above). The time interval is chosen
to allow a difference in domain size for late times of approximately 30
for the bulk fluid case, and 50 in the tortuous porous medium.  In both
cases this represents an increase in growth by an approximate factor of two.
Fig. 3 does not provide evidence to indicate the breakdown of dynamical
scaling in the porous medium, contrary to the results reported in
reference \cite{cha}.

The other geometries that we have studied show similar growth and
scaling behavior. We find
that the degree of slowing down is a function of the pore size
and porosity. Figure 4 shows the domain growth for two similar
geometries: one having periodic slits of radius 16 units ($\Diamond$ -
non-wet, {\small $\bigcirc$} - wet) and the other with 64 units
($\Box$ - non-wet,
$\triangle$ - wet).  Note here that, even for the wetting cases, the domain
size does not exceed the characteristic size of the porous medium, as seen in
Figure 2b.  This is because the confinement of the slit geometry
only allows for growth in one dimension, after some initial time.
In order to study slight deviations in the kinetics from the bulk case,
we considered geometry $(c)$ - a high-porosity medium (95\%) consisting
of a random distribution of disks.  In results not shown here, we
observed, for the non-wetting case, a slight slowdown in the kinetics,
as expected, since the
fluid experiences a certain degree of confinement. However, the
domain growth in the wetting case is indistinguishable from growth
of the bulk fluid.  Our simulations indicate that confinement of the
fluid within a pore geometry appears to be the key to the slowdown of
kinetics. The form of the geometry, whether random, tortuous and
interconnected or slit, does not appear to play a role in determining
if slowdown occurs or not.

In conclusion, we have used a lattice Boltzmann model to study the
kinetics of domain growth of a binary fluid in a number of two dimensional
geometries resembling porous media. Our simulation results, carried out for
large lattices, do not support the breakdown of late-time dynamical scaling,
indicating the absence of many length scales at late times. Our results
suggest that confinement of the fluid in a region leads to the slow
kinetics.  Moreover, they indicate that the random field model, which
requires some random component at a surface to obtain the slow kinetics,
may not be relevant to the problem.

We thank F. J. Alexander, G. D. Doolen and K. G. Eggert for
useful comments and suggestions. This work was supported by the
U. S. Department of Energy at Los Alamos National Laboratory.
Numerical simulations were carried out using the computational
resources at the Advanced Computing Laboratory at the Los Alamos
National Laboratory.

\newpage
\noindent {\bf Figure Captions}\par
\bigskip
\noindent {Figure 1.} Domain morphology at $t=4000$ for two geometries
showing $(a)$
wetting and $(b)$ non-wetting cases. Figure $(a)$ is a random distribution
of disks of radius 12 units, variance 4 units and porosity 95\%, and
Figure $(b)$ is an interconnected and tortuous geometry with porosity 75\%,
previously used in \cite{cha}. The rock is shown in white, and the two fluids
in black and grey. The wetting fluid for $(a)$ is black.  Note the elongated
domains with a layer of wetting fluid in $(a)$, and the characteristic plugs,
cusps, and straight interfaces of the non-wet condition in $(b)$.\par
\medskip

\noindent {Figure 2.} The variation of domain size as a function of
time for the geometry of Fig. 1(b). (a) For a $4096 \times 4096$ lattice,
the curves are for the bulk fluid ({\small $\bigcirc$}), wetting ($+$),
and non-wet ($\Diamond$) conditions. The dashed line at 160.8 represents the
characteristic size of the porous medium, and the dotted line
emphasizes the early ($t^{1/2}$) and late ($t^{2/3}$) growth of the bulk fluid.
(b) The effects of the slip condition ({\small $\bigcirc$}) on domain growth
versus two non-slip conditions: wet ($+$) and non-wet ($\Diamond$) ($1024
\times
1024)$. The characteristic pore size is shown with the dashed line
at 40.9. (c) A comparison of the growth for the red ($\Box$) and blue
({\small $\bigcirc$}) fluids in the presence of wetting $(1024 \times 1024$,
red is wetting). The domain growth for the red and blue fluids overlap for the
non-wetting case ($\triangle$) $(1024 \times 1024)$.\par
\medskip

\noindent {Figure 3.} The scaled structure factor $F(y)$ as a function
of $y\ (=kR(t))$ for $(a)$ the bulk fluid and $(b)$ the
non-wet tortuous geometry \cite{cha} for $4096 \times 4096$.  Scaling
was investigated, for the bulk fluid, over time interval $[1000,3000]$,
and, for the porous medium, over time interval $[7000,21000]$.  Both
of these time intervals represent respective domain growth of an
approximate factor of two.  A comparison of $(a)$ and $(b)$
makes it difficult to conclude that late-time dynamical scaling in the
porous medium breaks down, as reported in reference \cite{cha}.\par
\medskip

\noindent {Figure 4.}  The effects of pore size and wetting on the
domain growth for a slit geometry with pore radii 16 units ($\Diamond$
non-wet, {\small $\bigcirc$} wet, characteristic pore size 25.6) and 64 units
($\Box$ non-wet, $\triangle$ wet, characteristic pore size 101.3),
using a lattice size of $1024 \times 1024$.  Even for the wetting case,
the domain size does not exceed the characteristic size of the porous
medium.  This is because the confinement of the slit geometry
only allows for growth in one dimension, after some initial time.\par
\end{document}